
\input begla \centerline{\large ASYMPTOTIC CLASSIFICATION OF
SOLUTIONS}\par

\centerline{\large OF THE DISCRETE PAINLEV\'E-1 EQUATION} \vskip .8cm

\centerline{Vadim L. Vereschagin \footnote{Supported by ISF grant RK2000}}
\vskip 1.0cm \par

\noindent{\it Irkutsk Computing Center, Lermontov Str. 134, 664033 IRKUTSK,
RUSSIA} \vskip 1.cm \par

{\bf Abstract.} The main subject of the paper is the so-called Discrete
Painlev\'e-1 Equation (DP1). Solutions of the DP1 are classified under
criterion of their behavior while argument tends to infinity. The
appropriate theorems of existence are proved.  \vskip 1.cm \par

{\bf Introduction.} Recent rush of interest to ordinary differential
equations of the Painlev\'e type (see \cite{1} ) is provided mostly by two
reasons. The first one consists in rich history of the Painlev\'e equations
as a classical object of the ODE analytical theory. The second one is that
they arise in a number of concrete problems in different aspects of
theoretical and mathematical physics related to nonlinear evolutionary
equations, quantum field theory and statistical physics (see \cite{2,3,4}
). Analysis of some models of nonlinear theoretical physics indicates that
the Painlev\'e transcendents, describing self-similar regimes, play a role
very similar to one of classical special functions in linear problems. So,
the Painlev\'e-1 eq.(P1)

$$y''=6y^2+x\eqno(0.1)$$\noindent
describes waveforms of solutions of the Boussinesq eq., a role of P2

$$y''=xy+2y^3\eqno(0.2)$$\noindent
in asymptotic integration of nonlinear systems was investigated in papers
of Novokshenov, Its and others \cite{4,5} .

It is wellknown that the Painlev\'e transcendents are meromorphic functions
that cannot be expressed in terms of any known special functions (see
\cite{6} ). Therefore particular attention must be paid to asymptotic (as
$\vert x\vert\rightarrow\infty$) properties of the Painlev\'e eqs.  real
solutions (see \cite{7,8} ). Authors of paper \cite{9} developed an
asymptotic classification of the P1 eq.. Kapaev \cite{10} used the
isomonodromic deformations method and considerably advanced in this
direction. Novikov, Dubrovin and Krichever \cite{11,12} applied some
elements of finite-gap theory and Whitham method to the problem mentioned
above.

Studies in the theory of matrix model in two-dimensional gravity deal with
orthogonal polynomials with weight function

$$w(x)=exp[-V(x)],$$\noindent
where $V(x)$ is a polynomial in even powers. Equations

$$c^{1\over2}_{n}[V'(L)]_{n,n-1}=n \eqno(0.3)$$\noindent
where

$$L_{nm}=c^{1\over2}_{m}\delta_{n+1,m}+c^{1\over2}_{m}\delta_{n-1,m},\ n\in
{ \bf Z},$$\noindent
are usually referred to as the discrete string equations (see \cite{gross}
). Problem of description for asymptotic (as $n\rightarrow\infty$) behavior
of solutions $c_{n}$ of (0.3) was the subject of paper \cite{bauldry} . The
simplest nontrivial string equation, first obtained in \cite{bessis} ,
corresponds to a case $V(x)=g_{2}x^{2}+ g_{4}x^{4}$. Asymptotic (as
$n\rightarrow\infty$) formulae for $c_{n}$ while $g_{2}=0$ were written out
by Nevai \cite{nevai} . The main subject of this paper is to investigate
the general ($g_{2}\neq 0$) case, when the equation (0.3) takes the
following form:

$$c_{n}+4gc_{n}(c_{n-1}+c_{n}+c_{n+1})=\epsilon n+\nu;\ \
n=0,1,2,...,\eqno(0.4)$$\noindent
where $g,\ \nu,\ \epsilon>0$ are constant real-valued parameters, $c_{n}$
is an unknown sequence. There is also a procedure of transition to
continuum limit, transforming (4) into the Painlev\'e-1 equation (1):
$c_{n}=\rho [1-2ay(x)],\ where\ \epsilon^{-1} =Ba^{5/2},\ \epsilon
n=A(1+\delta a^2x),\ B=-72g,\ A=\rho /2,\ \delta =2/3,\ \rho =(24g)^{-1},\
under\ a\rightarrow 0$. Therefore the eq. (4) is usually referred to as
Discrete Painlev\'e-1 Equation (DP1). Moreover, the authors of \cite{13}
posed and solved in some different context a problem of isomonodromic
integration of (0.4) (see detailed description of the method in \cite{5}).
Since equation (0.4) seems not to have any solution expressable via known
special functions, only asymptotic (as $n\rightarrow \infty$) properties of
the solutions will be under survey.\par The main goal of this paper is to
construct asymptotic (as $n\rightarrow\infty$) classification of solutions
of (0.4).\par {\bf Asymptotic classification.} It's easy to see that scale
transformation  $c_{n}\rightarrow {{c_{n}}\over {12\vert g\vert}}$ can
reduce equation (0.4) to the form

$$c_{n}\pm {1\over 3}c_{n}(c_{n-1}+c_{n}+c_{n+1})=12\vert g\vert (\epsilon
n+\nu);\hskip.5cm n=0,1,2,...$$\noindent
It's enough to examine only case $12\vert g\vert\epsilon=1$.
(While assuming $12\vert g\vert\epsilon\ne 1$ one should change $n$ to
$12\vert g\vert\epsilon n$ in all the following formulas). Thus,

$$c_{n}\pm {1\over 3}c_{n}(c_{n-1}+c_{n}+c_{n+1})=n+\nu;\hskip.5cm
n=0,1,2,... \eqno(1.1^{\pm})$$\noindent
where $\nu\in [0,1]$. Since main properties of the equations $(1.1^{+}),
(1.1^{-})$ are similar and appropriate reasonings coinside, we shall
investigate only the model $(1.1^{+})$ and imply reference (1.1) as
$(1.1^{+})$. We classify solutions $c_{n}(\nu)$ of the initial (discrete)
system (1.1) in the following way. First we develop asymptotic
classification of solutions of pseudodifferential system

$$c(x)+4gc(x)[c(x-1)+c(x)+c(x+1)]={\epsilon}x,\hskip.3cm x\in{\Bbb
R}\eqno(1.1')$$\noindent
where $c(x)$ is a function of continuous variable. Share the set of all
solutions of (1.1$'$) into classes in accordance with principle to be
defined further. After having determined these classes of solutions of
system (1.1$'$) we restrict them on a lattice (i.e. obtain already
solutions of (1.1) and get the needed classification of solutions of the
initial difference system (1.1).\par Now we start to develop the asymptotic
($x\rightarrow\infty$) classification of solutions of the Cauchy problem
(1.1$'$), where i.d. are determined by continuous in $\nu\in [0,1]$
functions $c_{0}(\nu),\ c_{1}(\nu)$ so that

$$c_{1}(\nu)\ne0;\ c_{0}(1)=c_{1}(0);\ c_{1}(1)= {3\over
{c_{1}(0)}}-3- c_{1}(0)-c_{0}(0);\eqno(1.2)$$\noindent

$$c(x)=\left\{\matrix{c_{0}(x),\ x\in[0,1]\cr c_{1}(x-1),\
x\in(1,2]}\right } \eqno(1.3)$$\noindent

\par LEMMA 1.1. Solutions of the problem (1.1$'$-3) cannot have any
singularities other than finite order poles while $x\in{\Bbb R}$ is
finite.

\par Proof. One can easily express quantities $c(n+\nu)=c_{n}(\nu)$ via
$\nu,\ c_{0}(\nu),\ c_{1}(\nu)$ and represent $c_{n}(\nu)$ as a meromorphic
function of $\nu,\ c_{0}(\nu),\ c_{1}(\nu)$ and Since the functions $
c_{0}(\nu),\ c_{1}(\nu)$ are continuous in $\nu$, we obtain the claim for
$\nu\in (1,0)$. It left only to prove continuity in junctions $\nu=0,1$.
{}From (1.2) one concludes that $c_{0}(1)=c_{1}(0),\ c_{1}(1)=c_{2}(0)$. Use
the induction  procedure: let $c_{j}(0)=c_{j-1}(1),\ j=1,2,...,N;\ N\ge 2$.
Then

$$c_{N+1}(0)={{3N}\over {c_{N}(0)}}-3-c_{N}(0)-c_{N-1}(0)=$$\noindent

$${{3(N-1+1)}\over {c_{N-1}(1)}}-3-c_{N-1}(1)-c_{N-2}(1)=c_{N}(1)$$\noindent
Thus, solutions of the Cauchy problem (1.1$'$-3) are meromorphic functions
whose singularities can be only poles of finite order.  Therefore we must
differ the solutions according with their behavior in infinity
$x\sim\infty$ :

\par A$'$) solutions regular in infinity\par B$'$) solutions singular in
infinity with infinite number of poles in neighbourhood of infinity

\par The fact of A$'$-type solutions' existence is not obvious and needs a
special proof that will be presented $a\ posteriori$. Taking into account
importance of these solutions, we investigate them still supposing their
existence. Had restricted A$'$, B$'$-solutions of the problem  (1.1$'$-3)
on the lattice $x=n=1,2,...$, one can obtain solutions of (1.1) that
correspond to A, B-types respectively. We also demand to be able to invert
the transformation from continuous solutions of (1.1$'$-3) to discrete ones
of (1.1):

$$c(x)=c_{n}(\nu),\ x=n+\nu,\ n\in{\Bbb Z},\ \nu\in[0,1]$$\noindent
Introduce the following coordinates transform:

$$v_{0}(n)=c_{2n}c_{2n+1},\ v_{1}(n)=c_{2n+2}+c_{2n+1}+3,\eqno(1.4)$$\noindent
turning (1.1) to system

$$\left\{
\matrix{[v_{0}(n)-6n][v_{0}(n)-6n-3]=v_{0}(n)v_{1}(n)v_{1}(n-1)\cr
v_{0}(n)+v_{0}(n+1)=12n+9-v_{1}(n)[v_{1}(n)-3]}\right }\eqno(1.5)$$

\par LEMMA 1.2. If the A-type solutions of (1.1) exist, then they can be
obtained with inversion of formulas (1.4) only on set of solutions of (1.5)
with asymptotics

$$v_{0}(n)=O(n),\ n\rightarrow\infty$$\noindent
See proof of this and the following Lemmas in Appendix.\par

LEMMA 1.3. All the  A-type solutions of (1.1) can have only
asymptotics $c_{n}=O(\sqrt{n})$\par
Now we are ready to prove the following\par

THEOREM 1.1.Asymptotic (as $n\rightarrow\infty$) behavior of solutions of
(1.1) can part them only to three classes:\par

1) singular solutions\par

2) solutions with finite number of elements $\{c_{0}(\nu),\
   c_{1}(\nu),...,\ 0\}$\par

3) solutions with asymptotics $c_{n}=\pm\sqrt{n}+o(\sqrt{n})$\par

Proof. To prove the claim one needs to deal with solutions $c_{n}\sim
c_{0}(n) \sqrt{n}$, where $c_{0}(n)=O(1)$ (see Lemma 1.3). Substitution
into (1.1) results in nonlinear autonomous difference equation on function
$c_{0}(\nu)$:

$$c_{0}(n)[c_{0}(n-1)+c_{0}(n)+c_{0}(n+1)]=3\eqno(1.6)$$\noindent
Now we integrate the equation (1.6) within the class of functions of order
O(1). It is simple to verify that (1.6) can be represented in the following
way:

$$L_{n}A_{n}=A_{n+1}L_{n},$$ \noindent
where $L_{n},\ A_{n}$ are 2$\ast$2 matrices: $L_{n}=\pmatrix{\lambda &
c_{0}(n)\cr -1 & 0}$,

$$A_{n}=\pmatrix{-\lambda(\lambda^2+2c_{0}(n)) & -2c_{0}(n)(\lambda^2+
c_{0}(n)+c_{0}(n+1))\cr 2(\lambda^2+c_{0}(n)+c_{0}(n-1)) &
\lambda(\lambda^2+2c_{0}(n))}$$ \noindent
where $\lambda$ is an arbitrary parameter. Spectral characteristics of
matrix $A_{n}$ are integrals of the equation (1.6).

$$detA_{n}=\lambda^6 +4\lambda^2 c_{0}(n)[c_{0}(n-1)+c_{0}+(n)c_{0}(n+1)]$$

$$+4c_{0}(n)[c_{0}(n-1)+c_{0}(n)][c_{0}(n)+c_{0}(n+1)]$$\noindent
The last term of this polynomial in $\lambda$ is an integral of (1.6):

$$c_{0}(n)[c_{0}(n-1)+c_{0}(n)][c_{0}(n)+c_{0}(n+1)]=3[c_{0}(n)+c_{0}(n+1)]$$

$$-c_{0}(n)c_{0}(n+1)[c_{0}(n)+c_{0}(n+1)]=D\equiv const$$\noindent
Denoting $c_{0}(n)=x,\ c_{0}(n+1)=y$, one obtains the Riemann surface of
genus 1 $\Gamma(x,y)$:

$$x^2 y+y^2 x-3(x+y)+D=0\eqno(1.7)$$\noindent
Elliptic uniformization of $\Gamma(x,y)$ into the standard torus $T^2$ is
determined by functions $\footnote[1]{See standard notation of the
Weierstrass elliptic functions theory  in book \cite{14}.}$

$$x(z)=V[\zeta(z)-\zeta(z-U)+\zeta(U)-\zeta(2U)]$$

$$y(z)=V[\zeta(z+U)-\zeta(z)+\zeta(U)-\zeta(2U)]\eqno(1.8)$$\noindent
where $U,V$ are parameters, $z\in T^2$. The quantity $U$ is determined
by identity $2\zeta(U)-\zeta(2U)=0$ or, equivalently,

$$\wp''(U)=0,\eqno(1.9a)$$

$$V^2=3[\wp(2U)-\wp(U)]^{-1}\eqno(1.9b)$$\noindent
Formula for solution of the equation (1.4) ensues from (1.8):

$$c_{0}(n)=V[\zeta(Un+P)-\zeta(Un-U+P)-\zeta(U)]\eqno(1.10)$$\noindent
where $P$ is a parameter; $U,V$ are specified by conditions (1.9).  Roots
of the equation (1.9a) $U_{j},\ j=1,2,3,4$ are located in the parallelogram
of periods so:

$$U_{1}=\omega+u_{1},\ U_{2}=\omega'+u_{2},\ where\ iu_{1},\ u_{2}\in{\Bbb
R},$$

$$U_{3}=\omega-u_{1},\ U_{4}=\omega'-u_{2};\ 0<\vert u_{1}\vert <\vert
\omega'\vert,\ \ 0<\vert u_{2}\vert <\vert \omega\vert$$\noindent
In case of inparticular curve $\Gamma(x,y)$ (1.7) the function (1.10)
enjoys the condition $c_{0}(n)=O(1)$ only if $c_{0}(n)$ is periodic in $n:\
c_{0}(n+N)=c_{0}(n),\ n=0,1,2,...$.(else $c_{0}(n)$ is unbounded). But
solutions of (1.1) of this kind cannot lead to the A$'$-type solutions of
(1.1$'$-3) because substitution of nonintegral numbers instead of $n$ into
the formulae (1.10) results in nonreal expansion coefficients. Therefore
solutions of (1.1) with coefficients (1.10) at major term do not belong to
the type A in case of inparticular curve. Investigate now the case of
particular curve $\Gamma(x,y)$

$$\nabla\Gamma (x,y)=0\Rightarrow\left\{ \matrix{y^2 +2xy-3=0\cr
x^2 +2xy-3=0}\right\}\Rightarrow x^2 -y^2=0\eqno(1.11)$$\noindent
There are two subcases:\par

$a)x=-y\Rightarrow (x+y)(xy-3)=0$. The appropriate solution is

$$c_{0}(4n)=c,\ c_{0}(4n+1)=-c,\ c_{0}(4n+2)=-{3\over c},\ c_{0}(4n+3)=
{3\over c},\eqno(1.12)$$\noindent
where $c\in {\Bbb C}$ is an arbitrary parameter. Prove now that solutions
of (1.1) with coefficients at major term (1.12) do not exist. First
substitute the following quantities into (1.1):

$$c_{4n}=\sqrt{n}(c+b_{4n}),\ c_{4n+1}=\sqrt{n}(-c+b_{4n+1}),$$

$$c_{4n+2}=\sqrt{n}(-{3\over c}+b_{4n+2}),\ c_{4n+3}=\sqrt{n}({3\over
c}+b_{4n+3})$$ \noindent
and investigate the appropriate linear equations on $b_{n}=o(1)$:

$$A_{n}=\pmatrix{b_{4n-1}+\alpha b_{4n}+b_{4n+1}\cr b_{4n}+\alpha
b_{4n+1}+b_{4n+2}\cr b_{4n+1}+\beta b_{4n+2}+b_{4n+3}\cr b_{4n+2}+\beta
b_{4n+3}+b_{4n+4}},\ where\ \alpha=1+{3\over {c^2}},\ \beta=1+{{c^2}\over
3},\  A_{n}=O(n^{-1/2})$$\noindent
in case $(b_{n}n^{1/2})^{-1}=o(1)\ and\ A_{n}=-3(1,1,1,1)^{T}+O(n^{-1/2})$,
if $b_{n}=O(n^{-1/2})$. Solving these linear equations we reduce:
$b_{4n+4}=2b_{4n}-b_{4n-4}$, whence we conclude that quantities $\vert
b_{4n}\vert\neq 0$ linearly grow and $ b_{n}\neq o(1)$. So the only
subcase is \par $b)\ x=y=\pm 1,\Rightarrow c_{n}(\nu)=\pm\sqrt{n}+...$\par

LEMMA 1.4. If the system (1.1) has the A-type solutions then they are
monotonous starting from some large number $n$.\par To prove the claim one
needs only to compute a few terms of expansions for the solutions
$c_{n}^{\pm}(\nu)\sim\pm\sqrt{n}$:

$$c_{n}^{\pm}(\nu)=\pm\sqrt{n}\left(1+{{\nu+1/4}\over{2n}}\right)-{1\over
2}+ o(n^{-1/2})$$\noindent
Sequence $c_{n}^{+}(\nu),\ (c_{n}^{-}(\nu))$, obviously, increase
(decrease) in $n$.\par

Now we can start to prove a set of statements on existence of the solutions
mentioned above. Denote $c_{n}(\nu;c_{0}(\nu),c_{1}(\nu))$ solution of the
Cauchy problem (1.1) with initial data $c_{0}(\nu),\ c_{1}(\nu)$. First
express $c_{2}(\nu),\ c_{3}(\nu)$ via polynomials of $c_{0}(\nu),\
c_{1}(\nu)$.  Analysis of these polynomials yields the following: for any
$\nu\in[0,1]$ one can find regions $\Omega_{0},\ \Omega_{1}\subset {\Bbb
R}$ such that for any i.d.  $c_{j}(\nu)\in\Omega_{j},\ j=0,1$ the
appropriate solutions of (1.1) enjoy the order condition:

$$0<c_{0}(\nu)<c_{1}(\nu)<c_{2}(\nu)<c_{3}(\nu).$$

\par LEMMA 1.5. If for some solution $c_{n}$ of equation (1.1) and number
$N\ge 3$ condition $0<c_{0}<c_{1}<...<c_{N-1}=c_{N}$ holds then following
two conditions: $c_{N+1}>c_{N}$ and $c_{N+2}<c_{N+1}$ also hold.\par Proof.
Let $0<c_{0}<c_{1}<...<c_{N-1}=c_{N}$. Then

$$c_{N+1}={{3(N+\nu)}\over {c_{N}}}-2c_{N}-3,\ c_{N-2}={{3(N+\nu-1)}\over
{c_{N}}}-2c_{N}-3<c_{N}\Rightarrow $$

$$c_{N+1}={{3}\over {c_{N}}}+c_{N-2},\hskip.5cm c_{N}+{{3}\over
{c_{N}}}-c_{N+1}>0\eqno(1.13)$$

$$c_{N-3}={{3(N+\nu-2)}\over {c_{N-2}}}-c_{N}-c_{N-2}-3<c_{N-2}=
{{3(N+\nu-1)}\over {c_{N}}}-2c_{N}-3;$$\noindent
Using (1.13), we obtain:

$${{3c_{N}(N+\nu-1)}\over {c_{N}c_{N+1}-3}}-c_{N+1}+{6\over {c_{N}}}<
{{3(N+\nu)}\over {c_{N}}}-c_{N},$$
\noindent or, equivalently,

$$c_{N+1}-c_{N}>{{3(N+\nu-2)}\over {c_{N}c_{N+1}-3}}\left(c_{N}-c_{N+1}+
{3\over {c_{N}}}\right)>0.$$

To prove the second part of the Lemma we must express quantities $c_{N+1},\
c_{N+2}$ via $c_{N}$. Condition $10c_{N}^{2}<-10c_{N}+10(N+\nu)$ ensues
from $c_{N}<c_{N+1}$,

$$2c_{N}^{4}+5c_{N}^{3}-c_{N}^{2}[8(N+\nu)-12]-9c_{N}(N+\nu-1)+
6(N+\nu-1)^2>0$$\noindent
- from $c_{N-3}<c_{N-2}$
Thus, we have

$$2c_{N}^{4}+5c_{N}^{3}-c_{N}^{2}[8(N+\nu)-2]-9c_{N}(N+\nu)+6(N+\nu)^2>0,$$
\noindent or, equivalently, $c_{N+2}<c_{N+1}$\par

THEOREM 1.2. The equation (1.1) has two not less than one-dimensional
families of solutions, monotonous in $n$.\par

Proof. We fix the quantity $c_{0}(\nu)\in\Omega_{0}$ and look for a number
$c_{1} (\nu)$, corresponding to the monotonous solution (1.1)
$c_{n}(\nu;c_{0}(\nu), c_{1}(\nu))$. Denote $\Omega_{1}^{k}$ a region in
{\Bbb R} such that if $c_{1}(\nu)\in\Omega_{1}^{k}$ than solution of (1.1),
corresponding to i.d.  $c_{0}(\nu),\ c_{1}(\nu)$, enjoys the other
condition:  $0<c_{0}(\nu)<c_{1}(\nu)<...<c_{k}(\nu)$. We know that regions
$\Omega_{1}^{k} \subset {\Bbb R}$ exist while $k=2,3$. We shall construct
the sets $\Omega_{1}^{k}$ for arbitrary numbers $k=4,5,...$.\par

Lemma 1.5 means that the equation (1.1) cannot have solutions enjoying
condition

$$0<c_{0}<c_{1}<...<c_{k-1}=c_{k}<c_{k+1}<...<c_{N},\ where\ k<N-1,$$
\noindent
therefore points, bounding the set $\Omega_{1}^{N}$, correspond to
condition

$$0<c_{0}<c_{1}<...<c_{N-2}=c_{N-1}\ or\
0<c_{0}<c_{1}<...<c_{N-2}<c_{N-1}=c_{N}$$
\noindent
Elementary analysis of polynomials leads to fact of existence of quantities
$(c_{1}^{3})^{\pm}$ such that $0<c_{0}<(c_{1}^{3})^{-}=c_{2}(\nu;c_{0},
(c_{1}^{3})^{-})$ and
$0<c_{0}<(c_{1}^{3})^{+}<c_{2}(\nu;c_{0},(c_{1}^{3})^{+})
=c_{3}(\nu;c_{0},(c_{1}^{3})^{+})$. Condition $(c_{1}^{3})^{-}\neq
(c_{1}^{3})^{+}$ ensues from Lemma 1.5. Denote the interval
$(c_{1}^{3})^{-},(c_{1}^{3})^{+})= \Omega_{1}^{3}$. One can extract from
Lemma 1.5: $c_{3}(\nu;c_{0},(c_{1}^{3})^{-})-
c_{4}(\nu;c_{0},(c_{1}^{3})^{-})>0$ and $c_{3}(\nu;c_{0},(c_{1}^{3})^{+})-
c_{4}(\nu;c_{0},(c_{1}^{3})^{+})<0$. As $0<c_{0}<c_{1}\leq c_{2}\leq
c_{3}$, we operate within a region of continuity of function
$c_{3}(\nu;c_{0},c_{1})- c_{4}(\nu;c_{0},c_{1})$ in argument $c_{1}$. So,
there is a number $(c_{1}^{4})^{-}\in\Omega_{1}^{3}$ such that
$c_{3}(\nu;c_{0},(c_{1}^{4})^{-})- c_{4}(\nu;c_{0},(c_{1}^{4})^{-})=0$. If
there are a few such numbers, we choose the nearest to $(c_{1}^{3})^{+}$
one. Denote $(c_{1}^{4})^{+}=(c_{1}^{3})^{+}$ and
$\Omega_{1}^{4}=((c_{1}^{4})^{+},(c_{1}^{4})^{-})\subset\Omega_{1}^{3}$.
Similarly we find number $(c_{1}^{5})^{+}\in\Omega_{1}^{4}$ such that
$c_{4}(\nu;c_{0},(c_{1}^{5})^{+})=c_{5}(\nu;c_{0},(c_{1}^{5})^{+})$ and
interval
$\Omega_{1}^{5}=((c_{1}^{5})^{-},(c_{1}^{5})^{+})\subset\Omega_{1}^{4}$,
where $(c_{1}^{5})^{-}=(c_{1}^{4})^{-}$. In this way we obtain the interval
sequence $\Omega_{1}^{3}\subset\Omega_{1}^{4}\subset...
\subset\Omega_{1}^{N} \subset ...$, such that for any
$c_{1}\in\Omega_{1}^{N}$ the appropriate solution satisfies the condition
$0<c_{0}<c_{1}<...<c_{N}$. The Theorem for monotonously decreasing
solutions $(c_{n}^{-}\sim -\sqrt{n})$ can be proven in the similar way.

\par  Now we prove the following \par

LEMMA 1.6. All the monotonous solutions of (1.1) belong to two
one-dimensional families of solutions with asymptotics $\pm\sqrt{n}$.\par

Proof. One needs only show that dimension of family the monotonous
solutions mentioned above does not exceed one. Examine the Cauchy problem
(1.1) with i.d. in points $n_{0},\ n_{0}+1$, where $n_{0}$ is a large
number. Following Lemma 1.4., we have series

$$c_{n_{0}}=\sqrt{n_{0}}-{1\over 2}+{1\over 2}\left({1\over 4}+\nu\right)
n_{0}^{-1/2}+o(n_{0}^{-1/2})\eqno(1.14)$$
\noindent
such that there exists a number $c_{n_{0}+1}>c_{n_{0}}$, determined also by
the formula (1.14) (with $n_{0}+1$ instead of $n_{0}$), and solution of the
Cauchy problem (1.1) with the i.d. $c_{n_{0}},\ c_{n_{0}+1}$ is monotonous
and has asymptotics $\sqrt{n}$. Assume that for some fixed $c_{n_{0}}$
there exists an one-dimensional neighbourhood $\Delta$ of the point
$c_{n_{0}+1}$ enjoying the following condition: for any
$c'_{n_{0}+1}\in\Delta$ the appropriate solution of (1.1) with i.d.
($c_{n_{0}},\ c'_{n_{0}+1}$) is monotonous and, consequently, also has
asymptotics $\sqrt{n}$. Let $\delta_{1}=c'_{n_{0}+1}-c_{n_{0}+1},\
\vert\delta_{1}\vert\ll 1$. We shall compute solutions of slightly
perturbed Cauchy problem (1.1) with i.d. ($c_{n_{0}},\
c'_{n_{0}+1}=c_{n_{0}+1}+\delta_{1}$) and denote them $c_{n_{0}},\
c_{n_{0}+1}+\delta_{1},\ c'_{n_{0}+2}=c_{n_{0}+2}+ \delta_{2},...$. Simple
to verify that $\delta_{2}=-4\delta_{1}+O(\delta_{1} {n_{0}}^{-1/2})$,
$\delta_{3}=15\delta_{1}+O(\delta_{2}{n_{0}}^{-1/2})$,...,
$\delta_{N+1}=-4\delta_{N}-\delta_{N-1}+O(\delta_{N}N^{-1/2})$. The
sequence $\delta_{N}$ with alternating signs grows in module exponentially

$c'_{n_{0}+N+1}-c'_{n_{0}+N}=c_{n_{0}+N+1}-c_{n_{0}+N}+\delta_{N+1}-
\delta_{N}$.  As follows from Lemma 1.4, the monotonous solutions $c'_{n}$
must enjoy conditions $\vert
c'_{n_{0}+N+1}-c'_{n_{0}+N}\vert=O\left({1\over{\sqrt{n_{0}+N}}}\right)$.
But quantity $\vert\delta_{N+1}-\delta_{N}\vert$ gets not less than $\vert
c'_{n_{0}+N+1}-c'_{n_{0}+N}\vert$ sooner (in $N$) than terms of order
$O\left({{\delta_{N}}\over{\sqrt{n_{0}+N}}}\right)$ become innegligible in
the formulae above. In other words, in finite number of steps in $N$ the
perturbed solution $c'_{n}$ will cease to be monotonous.\par

Consequence. There are two and only two one-dimensional families of
monotonous solutions of the equation (1.1) with asymptotics $\pm\sqrt{n}$.

Our next goal is to prove that the classification, developed above, is
reasonable.  Thus, we must show existence of continuous solutions
(A$'$-type) of the pseudodifferential equation (1.1$'$). As mentioned
above, elementary analysis of some polynomials of order not more than four
yields the following. Let i.d. $(c_{0}(0),\ c_{1}(0)=c_{0}(1))$ lead to
monotonous solution of the problem (1.1). There is a region $\Omega\
((c_{0}(0),0),\ (c_{0}(1),1) \in\Omega)$ within a rectangle
$\{[c_{0}(0),c_{0}(1)]\times [0,1]\}$ such that any point
$(c_{0}(\nu),\nu)\in\Omega$, being taken with certain number $c_{1}
(\nu)>c_{0}(\nu)$ as i.d. of the problem (1.1), leads to solution,
satisfying the condition $0<c_{0}(\nu)<c_{1}(\nu)<c_{2}(\nu)<c_{3}(\nu)$

THEOREM 1.3. For any continuous function $c_{0}(\nu),\ \nu\in[0,1]$ such
that one can find a continuous function $c_{1}(\nu),\ \nu\in[0,1]$ such that
the conditions (1.2) hold and solution of the Cauchy problem  (1.1$'$-3) is a
continuous function with asymptotics $c(x)\sim\sqrt{n},\ x\rightarrow\infty$.

\par Proof. Theorem 1.2 indicates that for any point
$(c_{0}(\nu),\nu)\in\Omega$ one can find a number $c_{1}(\nu)$ so that
solution of the problem (1.1) with i.d.  $c_{0}(\nu),c_{1}(\nu)$ is
monotonous. We must show continuity in $\nu\in[0,1]$ of function,
constructed in this way. The point $c_{1}(\nu)$ is an unique common point
of the interval sequence $\Omega_{1}^{N},\ N\rightarrow\infty$ (see proof
of Theorem 1.2). Bounds of the regions $\Omega_{1}^{N}$ correspond to
solutions of (1.1), enjoying the condition $c_{0}(\nu)<c_{1}^{N}(\nu)<...<
c_{N-1}(\nu)=c_{N}(\nu)$. Using the condition $c_{N-1}(\nu)=c_{N}(\nu)$ as
an equation on $c_{1}^{N}(\nu)$, we conclude that the quantities
$c_{1}^{N}(\nu)$ are roots of polynomials with coefficients, smoothly
depending on $c_{0}(\nu)$ and $\nu$.  Analysing these polynomials and
utilizing Lemma 1.6, we derive that the sequence
$\{c_{1}^{N}(\nu)\}_{N=1}^{\infty}$ converges uniformly in $\nu\in[0,1]$.
Therefore the function $c_{1}(\nu)$ is continuous. Denote $c_{N}(\nu)=
c_{N}(\nu;\ c_{0}(\nu),c_{1}(\nu))$. Obviously, this function is partly
continuous in arguments $(\nu,c_{0},c_{1})$ and can have no singularities
but finite-order poles. Since the i.d. $c_{0}(\nu),c_{1}(\nu)$ are chosen
so that the appropriate quantity $c_{N}(\nu)=c_{N}(\nu;\
c_{0}(\nu),c_{1}(\nu))$ is finite for all $\nu\in [0,1]$, than $c_{N}(\nu)$
is also continuous in $\nu$. Now we construct solution of the equation
(1.1$'$): $c(x)=c_{[x]}(x-[x])$, where $[x]$ is the Entier function.
Utilizing Lemma 1.1, we obtain the claim.  Reasonings for the decreasing
solutions $c(x)\sim -\sqrt{x}$ are similar. \par

Conclusion. Now we can describe the asymptotic classification of solutions
of the equation (1.1): \par

A. There are two one-dimensional families of solutions, monotonous starting
from some number, with asymptotics $\sqrt{n}$ and $-\sqrt{n},\
n\rightarrow\infty$. We call these solutions regular in the following
sense: the A-type solutions can be obtained with restriction on a lattice
of solutions of the appropriate pseudodifferential solution (1.1$'$):
$c_{n}(\nu)=c(n+\nu),\ n=0,1,2,...;\ \nu\in[0,1)$ such that $c(x)$ is
regular in infinity.\par

B. Solutions singular in infinity. These solutions are the restriction of
singular solutions of the equation (1.1$'$).\par

REMARK. The case $g<0$, corresponding to the equation $(1.1^{-})$, leads to
classification, similar to one developed above where the regular (A-type)
solutions have asymptotics $\pm (-1)^{n}\sqrt{3n}$.
\par More precise asymptotics of the solutions will be calculated in the
next paper.\par

\vskip.6cm\par APPENDIX. Proof of Lemmas 1.2 and 1.3.\par

First prove the auxiliary \par

LEMMA A.1. Let

$$0<\alpha<2\Rightarrow \sum_{n=N_{0}>1}^{N}n^{\alpha}=
O(N^{\alpha +1})$$\noindent
as $N\rightarrow\infty$.\par

Proof. Denote

$$\Delta_{N}=\int_{N_{0}}^{N}x^{\alpha}dx-\sum_{n=N_{0}}^{N}n^{\alpha}=
\sum_{n=N_{0}}^{N-1}\int_{n}^{n+1}(x^{\alpha}-n^{\alpha})dx$$

$$=\sum_{n=N_{0}}^{N-1}\left[{{(n+1)^{\alpha +1}-n^{\alpha}}\over{\alpha
+1}} -n^{\alpha}\right]=\sum_{n=N_{0}}^{N-1}g_{n},$$
\noindent
where series

$$g_{n}={{\alpha}\over {2}}n^{\alpha-2}+{{\alpha(\alpha-1)}\over
{6}}n^{\alpha-3} +{{\alpha(\alpha-1)(\alpha-2)}\over
{24}}n^{\alpha-4}+...$$
\noindent
and series

$$g'_{n}={{\alpha}\over {2}}n^{\alpha-2}+{{\vert\alpha(\alpha-1)\vert}\over
{6}}n^{\alpha-3} +{{\vert\alpha(\alpha-1)(\alpha-2)\vert}\over
{24}}n^{\alpha-4}+...,$$
\noindent
evidently, converge; $n=2,3,...$

$$\vert g_{n}\vert<g'_{n}\hskip .8cm g'_{n+1}<g'_{n}\Rightarrow$$

$$\vert\Delta_{N}\vert=\vert\sum_{n=N_{0}}^{N-1}g_{n}\vert\leq\sum_{n=
N_{0}}^{N-1} \vert
g_{n}\vert<\sum_{n=N_{0}}^{N-1}g'_{n}<(N-1-N_{0})g'_{N_{0}}\Rightarrow$$
\noindent
the order of quantity $\vert\Delta_{N}\vert$ is not more than O(N) (we
denote this fact as $deg\vert\Delta_{N}\vert\leq 1)\Rightarrow$

$$0=\lim_{N\rightarrow\infty}{{\Delta_{N}}\over{N^{\alpha+1}}}={1\over{
\alpha+1}}
-\lim_{N\rightarrow\infty}{{1}\over{N^{\alpha+1}}}\sum_{n=N_{0}}^{N}
n^{\alpha}.$$

\par Proof of Lemma 1.2. First adopt notations. A sequence $\{x_n\}$ is to
be called of order $O(1)$ if it is bounded and does not converge to zero
while $n\rightarrow\infty$. The identity sign is realized in its asymptotic
sense:  $x_{n}\equiv a\Leftrightarrow$ one can find not more than finite
set of numbers $n_{1},n_{2},...,n_{m}$ such that $x_{n_{j}}\neq a$.\par

The main object of the Lemma is system

$$[v_{0}(n)-6n][v_{0}(n)-6n-3]=v_{0}(n)v_{1}(n-1)v_{1}(n)\eqno(I)$$

$$v_{0}(n)+v_{0}(n+1)=12n+9-v_{1}(n)[v_{1}(n)-3]\eqno(II)$$
\noindent
and its direct consequence - equation

$$v_{1}(n)[v_{1}(n-1)+v_{1}(n)+v_{1}(n+1)]=\frac{6n(6n+3)}{v_{0}(n)}+
\frac{(6n+6)(6n+9)}{v_{0}(n+1)}-12n-9\eqno(III)$$
\noindent
Notation $(I,n)$ will further mean considering of the equation (I) in point
$n$.  \par We shall look for asymptotics (as $n\rightarrow\infty$) of
solutions of the system (I-II) in form

$$v_{j}(n)=u_{j}(n)f_{j}(n)+...,\hskip .8cm j=0,1,\eqno(A2.1)$$
\noindent
where $u_{j}(n)=O(1),\ f_{j}(n)$ is monotonous. It is clear that $f_{0}(n)$
cannot satisfy $f_{0}(n)=o(n)$, else: $(II,n)\Rightarrow
v_{1}(n)=O(\sqrt{n}),\ (I,n)\Rightarrow$ left-hand part$\sim O(n^2)$,
right-hand part$\sim v_{0}(n)n \Rightarrow v_{0}(n)=O(n)!$.\par

Thus, it left to deal with solutions of (I-II) in the form (A.1), where the
function $f_{0}(n)$ is of order more than $n:\ {{n}\over {f_{0}(n)}}=o(1)$.
Devide the proof into two parts: the first one will deal with functions
$f_{0}(n)$ such that their moduli grow quicker than a linear function and

$$\lim_{n\rightarrow\infty}{{f_{0}(n)}\over{f_{0}(n+1)}}=1\eqno(i)$$\noindent
The second part deals with functions $f_{0}(n)$ such that

$${{f_{0}(n)}\over{f_{0}(n+1)}}=p_{n}>1\eqno(ii)$$
\noindent
starting from some number $n$.\par

{\bf Part One.} Introduce the following notation for the functions (i):
$f_{0}(n)\sim n^{x}$, where $x$ is formally an element from expansion
${\cal R} \supset {\Bbb R}$ of the set of real numbers. There are natural
relations among elements of the set ${\cal R}$:

$$x_{1}=x_{2}\Leftrightarrow f_{1}(n)/f_{2}(n)=O(1),\ where\ f_{j}(n)\sim
n^{x_{j}},\ j=1,2;$$

$$x_{1}>x_{2}\Leftrightarrow f_{1}(n)/f_{2}(n)=o(1),\ x_{1},x_{2}\in{\cal
R}$$
\noindent
Denote $degf(n)=x\Leftrightarrow f(n)\simn^{x},\ x\in{\cal R}$. Introduce a
linear structure in ${\cal R}$ over the field ${\Bbb R}$:

$$x_{1}+x_{2}=x_{3}\Leftrightarrow f_{3}(n)/[f_{1}(n)f_{2}(n)]=O(1),$$

$$x_{1}=-x_{2}\Leftrightarrow f_{1}(n)f_{2}(n)=O(1),$$

$$x_{1}=kx_{2},\ k\in{\cal R}\Leftrightarrow
f_{2}(n)/[f_{1}(n)]^{k}=O(1).$$
\noindent
Thus, let $v_{0}(n)=u_{0}(n)n^{a}+...$, where $a>1,\ a\in{\cal R},\
u_{0}(n)\neq 0.\ (I,n)\Rightarrow
[u_{0}(n)]^{2}n^{2a}=u_{0}(n)n^{a}v_{1}(n-1) v_{1}(n)$ $\Rightarrow
v_{1}(n)\sim n^{x},\ v_{1}(n)\sim n^{a-x}$ . If $x\neq a/2$, then we obtain
a contradiction with (II). Therefore, $v_{1}(n)=u_{1}(n)
n^{a/2}+...\Rightarrow$derive from (I): $u_{0}(n)=u_{1}(n-1)u_{1}(n),\
u_{1}(n) \neq 0$. Substituting into (II), we have in major order:

$$u_{1}(n)[u_{1}(n-1)+u_{1}(n)+u_{1}(n+1)]=0$$
\noindent
The general solution of this equation is

$$u_{1}(3n)=c_{0},\ u_{1}(3n+1)=c_{1},\
u_{1}(3n+2)=-c_{0}-c_{1},\eqno(A.2)$$
 \noindent
where $c_{0},\ c_{1}\in{\Bbb R}$ are arbitrary constants$\Rightarrow$

$$u_{0}(3n)=-c_{0}(c_{0}+c_{1}),\ u_{0}(3n+1)=c_{0}c_{1},\
u_{0}(3n+2)=-c_{1}( c_{0}+c_{1})$$
\noindent
Now we start successive examination of appropriate variants. Subcases are
denoted with additional letters\par

a) Let $c_{0}\neq 0,\ c_{1}\neq 0,\ c_{0}+c_{1}\neq 0\Rightarrow$ the
   right-hand part of (III,n) equals $-12n+o(n)$. Since $v_{1}(n)\sim
   n^{a/2}$, then

$$\left\{\matrix{v_{1}(3n-1)+v_{1}(3n)+v_{1}(3n+1)=-{{12}\over{c_{0}}}
n^{1-a/2}+... \cr
v_{1}(3n)+v_{1}(3n+1)+v_{1}(3n+2)=-{{12}\over{c_{1}}}n^{1-a/2}+...\cr
v_{1}(3n+1)+v_{1}(3n+2)+v_{1}(3n+3)={{12}\over{c_{0}+c_{1}}}n^{1-a/2}+
...}\eqno(A.3)$$
\noindent
whereas $1-a/2\geq 0$, else $3u_{1}(n)\equiv -12$, what contradicts with
(A.2) $\Rightarrow 1<a\leq 2$. Consider subcases.\par\hskip 2.cm

a1) Let $1<a<2$. Then, asymptotically integrating the linear system (A.3)
and using Lemma A.1, we obtain: $v_{1}(n)\sim n^{2-a/2}$, but
$2-{{a}\over{2}}> {{a}\over{2}}$, and get a contradiction with initial
supposition $v_{1}(n)\sim n^{a/2}$.  \par\hskip 2.cm a2) $a=2$: the system
(A.3) looks so:

$$\left\{\matrix{u_{1}(3n-1)+u_{1}(3n)+u_{1}(3n+1)=3-{{12}\over{c_{0}}} \cr
u_{1}(3n)+u_{1}(3n+1)+u_{1}(3n+2)=3-{{12}\over{c_{1}}}\cr
u_{1}(3n+1)+u_{1}(3n+2)+u_{1}(3n+3)=3+{{12}\over{c_{0}+c_{1}}}}$$
\noindent
Taking into account the equations (A.2), we obtain a system that leads to a
quadratic equation on either of the quantities $c_{0},\ c_{1}$ with
negative discriminant. To sum up the item a) we conclude: if $c_{0}\neq 0,\
c_{1}\neq 0,\ c_{0}+c_{1}\neq 0$, then the system (I-II) has no solutions
in the form (A.1-2).\par

b) Let $c_{0}=0,\ c_{1}\neq 0$. Then (see (A.2)):

$$u_{1}(3n)=0,\ u_{1}(3n+1)=c_{1},\ u_{1}(3n+2)=-c_{1}$$

$$u_{0}(3n)=0,\ u_{0}(3n+1)=0,\ u_{0}(3n+2)=-c_{1}^{2}$$
\noindent
Assume $v_{0}(3n)=-u_{0}^{c}(n)n^{c}+...$, where $c<a,\ u_{0}^{c}(n)\neq
0$.  Examine the appropriate variants \par\hskip 2.cm

b1) Let $1<c<a$. Then:

$$v_{0}(3n)=-u_{0}^{c}(n)n^{c}+...,\ \ v_{1}(3n+1)=c_{1}n^{a/2}+...,$$

$$v_{0}(3n+2)=-c_{1}^{a}n^{a}+...,\ \ v_{1}(3n+2)=-c_{1}n^{a/2}+...,$$
\noindent
Substitute these equalities into (I,3n):

$$u_{0}^{c}(n)n^{c}=-c_{1}(n-1)^{a/2}v_{1}(3n)+...\Rightarrow v_{1}(3n)=
-\frac{u_{0}^{c}(n)}{c_{1}}n^{c-a/2}+...\eqno(A.4)$$

$$(II,3n):\
u_{0}^{c}(n)n^{c}+v_{0}(3n+1)+...=12n-\frac{u_{0}^{c}(n)}{c_{1}}n^{c-a/2}
\left[\frac{u_{0}^{c}(n)}{c_{1}}n^{c-a/2}+3\right]$$

$$c>max\{1,2c-a,c-a/2\}\Rightarrow v_{0}(3n+1)=-u_{0}^{c}(n)n^{c}+...$$

$$(III,3n+1):\
v_{1}(3n)+v_{1}(3n+1)+v_{1}(3n+2)=-{{12}\over{c_{1}}}n^{1-a/2}+...\eqno(A.5)$$

$$(III,3n+2):\
v_{1}(3n+1)+v_{1}(3n+2)+v_{1}(3n+3)={{12}\over{c_{1}}}n^{1-a/2}+...,$$
\noindent
where $1-a/2\geq 0$, else $3c_{1}=-12=-3c_{1}$. Thus, $1<a\leq 2$. Using
(A.4,5) and Lemma A.1, we get: $c-a/2=2-a/2\Rightarrow c=2\Rightarrow a\leq
c$, what contradicts with the initial assumption of subcase b1).\par\hskip
2.cm

b2) Let $1=c<a$:

$$v_{0}(3n)=u_{0}^{1}(n)n+...,\ (u_{0}^{1}(n)\neq 0),\
v_{0}(3n+1)=c_{1}n^{a/2}+...,$$

$$v_{0}(3n+2)=-c_{1}^{2}n^{a}+...,,\ v_{1}(3n+2)=-c_{1}n^{a/2}+...$$

$$(I,3n):\
[u_{0}^{1}(n)-6]^{2}n^{2}=u_{0}^{1}(n)n(-c_{1})(n-1)^{a/2}v_{0}(3n)+...$$

$$\Rightarrow
v_{1}(3n)=-\frac{[u_{0}^{1}(n)-6]^{2}n^{1-a/2}}{c_{1}u_{0}^{1}(n)}+...
\eqno(A.6)$$

$$(II,3n): nu_{0}^{1}(n)+v_{0}(3n+1)=12n+...\ (as\ deg
v_{1}(3n)=1-a/2<1/2)$$

$$\Rightarrow v_{0}(3n+1)=[12-u_{0}^{1}(n)]n+...$$

\par\hskip 3.cm b2$\alpha)u_{0}^{1}(n)\neq 12$.\par

If $u_{0}^{1}(n)\neq 6$, then
$\frac{36}{u_{0}^{1}(n)}+\frac{36}{12-u_{0}^{1}(n)}-12\neq 0$.  Considering
the equations $(III,3n+1,3n+2)$, we obtain a contradiction, similar to one
reduced from (A.5). If $u_{0}^{1}(n)\equiv 6$, then:

$$(III,3n+1,3n+2):\left\{\matrix{v_{1}(3n)+v_{1}(3n+1)+v_{1}(3n+2)=-\frac{
6}{c_{1}}n^{1-a/2}+...  \cr
v_{1}(3n+1)+v_{1}(3n+2)+v_{1}(3n+3)=\frac{6}{c_{1}}n^{1-a/2}+...},$$
\noindent
whereas, likewise (A.5), conclude: $1<a\leq 2$. Substracting the equations,
we reduce: $\Delta_{n}v_{1}(3n)=\frac{12}{c_{1}}n^{1-a/2}$, but (A.6) means
$v_{1}(3n)=o(n^{1-a/2})$!

\par\hskip 3.cm b2$\beta)u_{0}^{1}(n)\equiv 12\Rightarrow v_{0}(3n+1)=o(n)$

$$(III,3n+1):v_{1}(3n)+v_{1}(3n+1)+v_{1}(3n+2)=$$

$$\frac{1}{c_{1}}n^{a/2}\left[\frac{(6n+6)(6n+9)}{v_{0}(3n+1)}-12n+...
\right]$$

$$(III,3n+2):v_{1}(3n+1)+v_{1}(3n+2)+v_{1}(3n+3)=-\frac{1}{c_{1}}n^{a/2}
(-9a)+...$$
\noindent
$\Rightarrow deg(\Delta_{n}v_{1}(3n))>1-a/2$, what contradicts with
(A.6).\par

c) Let $c<1<a$:

$$v_{0}(3n+2)=-c_{1}^{2}n^{a}+...,\ \ v_{1}(3n+1)=c_{1}n^{a/2}+...,$$

$$v_{0}(3n)=u_{0}^{c}(n)n^{c}+...,\ \ v_{1}(3n+2)=-c_{1}n^{a/2}+...,$$

$$(I,3n):36n^{2}=-u_{0}^{c}(n)n^{c}v_{1}(3n)c_{1}(n-1)^{a/2}+...
\Rightarrow$$

$$v_{1}(3n)=-\frac{36}{c_{1}u_{0}^{c}(n)}n^{2-c-a/2}$$

Let $v_{0}(3n+1)=u_{0}^{1}(n)n^{x}+...$. Consider the variants.\par

c1)$x>1$

$$(I,3n+1):[u_{0}^{1}(n)n^{x}-6n]^{2}=-u_{0}^{1}(n)n^{x}c_{1}n^{a/2}
\frac{36} {c_{1}u_{0}^{c}(n)}n^{2-c-a/2}+...\Rightarrow$$

$$u_{0}^{1}(n)n^{x}=\frac{36}{c_{1}u_{0}^{c}(n)}n^{2-c}+...\Rightarrow
x=2-c>1>c$$

$$(II,3n):u_{0}^{c}(n)n^{c}+u_{0}^{1}(n)n^{2-c}=-\left[\frac{36}{c_{1}
u_{0}^{c}(n)} \right]^{2}n^{4-2c-a}+...$$

$$\Rightarrow a+c=2\Rightarrow a=2-c=x!$$

c2)$x=1$

$$(I,3n+1):[u_{0}^{1}(n)-6n]^{2}n^{2}=-u_{0}^{1}(n)nc_{1}n^{a/2} \frac{36}
{c_{1}u_{0}^{c}(n)}n^{2-c-a/2}+...$$

$$\Rightarrow 3-c\leq 2\Rightarrow c\geq 1!$$

c3)$x<1$:

$$v_{0}(3n+2)=-c_{1}^{2}n^{a}+...,\ \ v_{1}(3n+1)=c_{1}n^{a/2}+...,$$

$$v_{0}(3n)=u_{0}^{c}(n)n^{c}+...,\ \ v_{1}(3n+2)=-c_{1}n^{a/2}+...,$$

$$v_{0}(3n+1)=u_{0}^{1}(n)n^{x}+...$$

$$(I,3n):36n^{2}=-c_{1}u_{0}^{c}(n)n^{c}v_{1}(3n)(n-1)^{a/2}+...
\Rightarrow$$

$$v_{1}(3n)=-\frac{36}{c_{1}u_{0}^{c}(n)}n^{2-c-a/2}\eqno(A.7)$$

$$(I,3n+1):36n^{2}=-c_{1}u_{0}^{c}(n)n^{x}\frac{36}{c_{1}u_{0}^{c}
(n)}n^{2-c-a/2} n^{a/2}+...\Rightarrow$$

$$u_{0}^{1}(n)=-u_{0}^{c}(n),\ \ x=c.$$

$$(II,3n):o(n)=12n-v_{1}(3n)[v_{1}(3n)-3]\Rightarrow v_{1}(3n)=\pm
2\sqrt{3n}+...  \Rightarrow$$

$$(see\ (A.7)):-\frac{36}{c_{1}u_{0}^{c}(n)}=\pm 2\sqrt{3},\
1/2=2-c-a/2\Rightarrow$$

$$u_{0}^{c}(n)=\mp \frac{6\sqrt{3}}{c_{1}},\ \ a+2c=3\Rightarrow$$
\noindent
We have the following asymptotics:

$$\left\{\matrix{v_{0}(3n+2)=-c_{1}^{2}n^{1+2\delta}+...,\ \ v_{1}(3n+2)=
c_{1}n^{1+\delta}+...\cr v_{0}(3n+1)=\pm
\frac{6\sqrt{3}}{c_{1}}n^{1-\delta}+...,\ \ v_{1}(3n+1)=
-c_{1}n^{1+\delta}+...\cr v_{0}(3n)=\mp
\frac{6\sqrt{3}}{c_{1}}n^{1+2\delta}+...,\ \ v_{1}(3n)=
\pm2\sqrt{3n}+...},$$
\noindent
where $\delta=1-c>0$.\par

Statement. Solutions of the system (I-II) with asymptotics (A.8) can yield
(with invertion of the formulae (1.4)) only B-type solutions of the
equation (1.1).\par

Proof. Suppose that solutions of (1.1) have regular asymptotics:

$$c_{0}(3n)\sim O(n^{x_{0}}),\ c_{0}(3n+1)\sim O(n^{x_{1}}),\
c_{0}(3n+2)\sim O(n^{x_{2}}),$$
\noindent
where $c_{0}(n)=c_{2n},\ c_{1}(n)=c_{2n+1}$. Then (see (1.4, A.8):

$$c_{1}(3n)\sim O(n^{1-\delta-x_{0}}),\ c_{1}(3n+1)\sim
O(n^{1-\delta-x_{1}}),\ c_{1}(3n+2)\sim O(n^{1+2\delta-x_{2}}),$$

$$max(x_{1},1-\delta-x_{0})\geq 1/2,\ \ max(x_{2},1-\delta-x_{1})\geq
1+\delta .$$
\noindent
Possible cases:\par

1)$x_{1}\geq 1/2 \Rightarrow x_{2}=1+\delta \Rightarrow c_{0}(3n+2)\sim
O(n^{1+\delta})$,\par $c_{1}(3n+1)\sim O(n^{1/2-\delta}),\ c_{1}(3n+2)\sim
O(n^{\delta}) \Rightarrow$

$$c_{0}(3n+2)[c_{1}(3n+1)+c_{0}(3n+2)+c_{1}(3n+2)+3]=O(n^{2+\delta}),$$
\noindent
what contradicts with (1.1);\par

2)$x_{1}<1/2\Rightarrow x_{0}=1/2-\delta\Rightarrow c_{1}(3n)=O(n^{1/2})$,

$$c_{1}(3n+1)=O(n^{1-x_{1}-\delta}),\ c_{0}(3n+1)=O(n^{x_{1}})\Rightarrow$$

$$deg\{c_{0}(3n+1)[c_{1}(3n)+c_{0}(3n+1)+c_{1}(3n+1)+3]\}=max(x_{1}+1/2,
1-\delta),$$
\noindent
what also contradicts with (1.1). The part One of the proof is over.\par

{\bf Part Two.}\par

The main objects of investigation here are possible solutions of the system
(I-II): $v_{0}(n)\sim f_{n}$, where $f_{n}$, is a function of (ii)-type:
$f_{n}/f_{n-1}=p_{n},\ p_{n}>1$, starting from some number $n$.

$$(I,n):v_{1}(n)=O(\sqrt{f_{n}})$$

$$(III,n,n+1):\matrix{v_{1}(n-1)+v_{1}(n)+v_{1}(n+1)=3+O\left(\frac{n^{2}}
{\sqrt{f_{n}}}\right)\cr
v_{1}(n)+v_{1}(n+1)+v_{1}(n+2)=3+O\left(\frac{n^{2}}{\sqrt{f_{n}}}\right)}$$

$$\Rightarrow
\Delta_{n}=v_{1}(3n+j)=3+O\left(\frac{n^{2}}{\sqrt{f_{n}}}\right),\
j=0,1,2$$
\noindent
Integrating this equation, we obtain: $v_{1}(3n+j)=O(1),\ j=0,1,2,$ what
contradicts with (I,n). Therefore, the series $\{v_{0}(n)\}$ contains
subsequences with asymptotics of order less than  $f_{n}$. Moreover, we can
write out:

$$\matrix{v_{0}(3n+2)=-c_{1}^{2}f_{n}+...,\ \
v_{1}(3n+2)=-c_{1}\sqrt{f_{n}}+...,\cr v_{0}(3n+1)=o(f_{n-1}),\ \
v_{1}(3n+1)=c_{1}\sqrt{f_{n}}+...,\cr v_{0}(3n)=o(f_{n-2})}\eqno(A.9)$$
\noindent
Examine the appropriate variants.\par

1) Let $degv_{0}(3n)>1$,

$$(I,3n):v_{1}(3n)=-\frac{v_{0}(3n)}{c_{1}\sqrt{f_{n}}}+...=o(\sqrt{
f_{n}}),$$
\noindent
else one gets contradiction with (II,3n):

$$v_{0}(3n)+v_{0}(3n+1)=12n+\frac{[v_{0}(3n)]^{2}}{c_{1}f_{n-1}}+...
\Rightarrow$$

$$1+\frac{v_{0}(3n+1)}{v_{0}(3n)}=\frac{v_{0}(3n)}{c_{1}^{2}f_{n-1}} +...=
\frac{o(f_{n-2})}{c_{1}^{2}f_{n-1}}+o(1)=o(1)\Rightarrow$$

$$\lim_{n\rightarrow\infty}\frac{v_{0}(3n+1)}{v_{0}(3n)}=-1\Rightarrow
degv_{0}(3n+1)>1$$

$$(III,3n+1,3n+2):\matrix{v_{1}(3n)+v_{1}(3n+1)+v_{1}(3n+2)=3-\frac{12n}
{c_{1}\sqrt{f_{n}}}+...  \cr
v_{1}(3n+1)+v_{1}(3n+2)+v_{1}(3n+3)=3+\frac{12n}{c_{1}\sqrt{f_{n}}}+...}$$

$$\Rightarrow
\Delta_{n}v_{1}(3n)=\frac{24n}{c_{1}\sqrt{f_{n}}}+...\eqno(A.10)$$

$$(III,3n):v_{1}(3n)c_{1}[\sqrt{f_{n}}-\sqrt{f_{n-1}}]=-12n+...
\Rightarrow$$

$$v_{1}(3n)=-\frac{12n}{c_{1}\sqrt{f_{n-1}}(\sqrt{p_{n}}-1)}+...\eqno(A.11)
\Rightarrow$$

$$v_{0}(3n)=12n(\sqrt{p_{n}}-1)+...\Rightarrow degp_{n}>0\ (as\
degv_{0}(3n)>1) \Rightarrow$$
\noindent
We have contradiction between the formulae (A.10 and 11).\par

2) Let $v_{0}(3n)=u_{0}(n)n+...$, where $u_{0}(n)\neq 0$

$$(I,3n):v_{1}(3n)=-\frac{n[u_{0}(n)-6]^{2}}{u_{0}(n)c_{1}\sqrt{f_{n-1}}}
+...  \eqno(A.12)$$

$$(II,3n):u_{0}(n)n+v_{0}(3n+1)=12n+...\Rightarrow
v_{0}(3n+)=[12-u_{0}(n)]n+...$$

\par $u_{0}(n)\neq 12.\ (III,3n+1,3n+2)$:

$$v_{1}(3n)+v_{1}(3n+1)+v_{1}(3n+2)=3+\frac{n}{c_{1}\sqrt{f_{n}}}\left[
\frac{36}{12-u_{0}(n)}-12\right]+...$$

$$v_{1}(3n+1)+v_{1}(3n+2)+v_{1}(3n+2)=3+\frac{n}{c_{1}\sqrt{f_{n}}}
\left[\frac{36} {u_{0}(n+1)}+12\right]+...$$

$\Rightarrow v_{1}(3n)\leq 0$

$$\Delta_{n}v_{1}(3n)=\frac{n}{c_{1}\sqrt{f_{n}}}\left[\frac{36}{u_{0}
(n+1)}+\frac{36}{12-u_{0}(n)}+24\right]+...\eqno(A.13)$$

$$(III,3n+1):v_{1}(3n)=\left[\frac{36}{u_{0}(n)}+\frac{36}{12-u_{0}(n)}-
12\right] \frac{n}{c_{1}\sqrt{f_{n-1}}(\sqrt{p_{n}}-1)}+...$$

$$(I,3n+1):[6-u_{0}(n)]^{2}=[12-u_{0}(n)]\left[\frac{36}{u_{0}(n)}+
\frac{36}{12-u_{0}(n)}-12\right] \frac{\sqrt{p_{n}}}{\sqrt{p_{n}}-1}+...$$

$$(I,3n):[6-u_{0}(n)]^{2}=-u_{0}(n)\left[\frac{36}{u_{0}(n)}+\frac{36}
{12-u_{0}(n)}-12\right] \frac{1}{\sqrt{p_{n}}-1}+...$$

$$\Rightarrow \sqrt{p_{n}}=\frac{u_{0}(n)}{u_{0}(n)-12}=O(1)\Rightarrow
u_{0}(n) >6\ (else\ \vert p_{n}\vert\leq 1)\eqno(A.14)$$
\noindent
and derive from (A.12):

$$\Delta_{n}v_{1}(3n)=\frac{n}{c_{1}\sqrt{f_{n}}}\left[\frac{[6-u_{0}
(n)]^{2}} {u_{0}(n)-12}-\frac{[6-u_{0}(n+1)]^{2}}{u_{0}(n+1)}\right]$$
\noindent
from (A.13):

$$\Delta_{n}v_{1}(3n)=\frac{n}{c_{1}\sqrt{f_{n}}}\left[\frac{36}
{u_{0}(n+1)}+\frac{36}{u_{0}(n)-12}+24\right]$$

$\Rightarrow u_{0}(n+1)-u_{0}(n)=-12-\frac{72}{u_{0}(n+1)}$, what
contradicts with (1.14) and condition $u_{0}(n)=O(1)$.\par

2b)$u_{0}(n)\equiv 12:\ v_{0}(3n)=12n+...$.

$$(I,3n): v_{1}(3n)=-\frac{3n}{c_{1}\sqrt{f_{n-1}}}+...;\ \
(II,3n):v_{0}(3n+1)=o(n);$$

$$(I,3N+1):v_{0}(3n+1)=-\frac{12n}{\sqrt{p_{0}}}+...\Rightarrow
degp_{0}>0.$$
\noindent
Summing up, we write out the asymptotics:

$$\matrix{v_{0}(3n)=12n+...,\ \
v_{1}(3n)=-\frac{3n}{c_{1}\sqrt{f_{n-1}}}+...,\cr
v_{0}(3n+1)=-\frac{12n}{\sqrt{p_{0}}}+...,\ \
v_{1}(3n+1)=c_{1}\sqrt{f_{n-1}}+...,\cr v_{0}(3n+2)=-c_{1}^{2}f_{n}+...,\ \
v_{1}(3n+2)=-c_{1}\sqrt{f_{n}}+...}\eqno(A.15)$$
\noindent
The system (A.15), being rewritten in terms of coordinates
$\{c_{0}(n),c_{1}(n)\}$, has the following form:

$$\matrix{c_{0}(3n)=\frac{6n}{c_{1}\sqrt{f_{n-1}}}+...,\ \
c_{1}(3n)=2c_{1}\sqrt{f_{n-1}}+...,\cr
c_{0}(3n+1)=-2{c_{1}\sqrt{f_{n-1}}}+...,\ \
c_{1}(3n+1)=\frac{6n}{c_{1}\sqrt{f_{n}}}+...,\cr
c_{0}(3n+2)={c_{1}\sqrt{f_{n}}}+...,\ \
c_{1}(3n+2)=-c_{1}\sqrt{f_{n}}+...}\eqno(A.16)$$
\noindent
Change the cordinates again:

$$w_{0}(n)=c_{1}(n)c_{0}(n+1),\ \ w_{1}(n)=3+c_{0}(n+1)+c_{1}(n+1)$$
\noindent
and obtain a system on $(w_{0},w_{1})$:

$$[w_{0}(n)-6n-3][w_{0}(n)-6n-6]=w_{0}(n)w_{1}(n-1)w_{1}(n)$$

$$w_{0}(n)+w_{0}(n+1)=12n+15-w_{1}(n)[w_{1}(n)-3],$$
\noindent
which differs from the system (I-II) only in argument shift
$n\rightarrown+1/2$.  Writing out the asymptotics (A.16) in coordinates
$(w_{0},w_{1})$, we get:

$$\matrix{w_{0}(3n)={-4n}{c_{1}^{2}{f_{n-1}}}+...,\ \
w_{1}(3n)=2c_{1}\sqrt{f_{n-1}}+...,\cr w_{0}(3n+1)=6n+...,\ \
w_{1}(3n+1)=-2c_{1}\sqrt{f_{n-1}}+...,\cr w_{0}(3n+2)=-6n+...,\ \
w_{1}(3n+2)=o\left(\frac{n}{\sqrt{f_{n}}}\right)+...}$$
\noindent
One can easily derive from 2a) that these solutions do not satisfy the
claim.\par

3)$degv_{0}(3n)<1:$\par

$$\matrix{v_{0}(3n+2)=-{c_{1}^{2}{f_{n}}}+...,\ \
v_{1}(3n+2)=-c_{1}\sqrt{f_{n}}+...,\cr v_{0}(3n+1)=o(f_{n-1}),\ \
v_{1}(3n+1)=c_{1}\sqrt{f_{n}}+...,\cr v_{0}(3n)=o(n)$$

$$(I,3n):v_{1}(3n)=-\frac{36n^{2}}{c_{1}\sqrt{f_{n-1}}v_{0}(3n)}+...$$\par

3a)$degv_{0}(3n+1)>1:$\par

$$(I,3n+1):v_{0}(3n+1)=-\frac{36n^{2}\sqrt{p_{n}}}{v_{0}(3n)}+...$$\par

$$(II,3n):-v_{1}^{2}(3n)=v_{0}(3n+1)+...\Rightarrow v_{0}(3n+1)=
-\frac{36n^{2}\sqrt{p_{n}}}{v_{0}(3n)}+...\Rightarrow$$

$$degv_{0}(3n+1)>deg(\sqrt{nf_{n}}),$$
\noindent
what contradicts with the assumption $v_{0}(3n+1)=o(f_{n-1})$.\par

3b)$v_{0}(3n+1)=u_{0}^{1}(n)n+...,\ \ u_{0}^{1}(n)\neq 0$

$$(I,3n+1):v_{1}(3n)=\frac{[u_{0}^{1}(n)-6]^{2}n}{c_{1}\sqrt{f_{n}}
u_{0}^{1}(n)}+...$$

$$(II,3n):u_{0}^{1}(n)\equiv 12\Rightarrow
v_{1}(3n)=\frac{3n}{c_{1}\sqrt{f_{n}}}+...=
-\frac{36n^{2}}{c_{1}\sqrt{f_{n-1}}v_{0}(3n)}+...$$

$$\Rightarrow v_{0}(3n)=-12n\sqrt{p_{1}}+...$$
\noindent
- contradiction with the assumption $v_{0}(3n)=o(n)$\par

3c)$v_{0}(3n+1)=o(n)$\par

$$(I,3n+1):36n^{2}=v_{0}(3n+1)v_{0}(3n)c_{1}\sqrt{f_{n}}+...$$

$$(II,3n):v_{1}(3n)=\pm 2\sqrt{3n}+...\Rightarrow v_{0}(3n+1)=\pm
\frac{6\sqrt{3n}}{c_{1}\sqrt{f_{n}}}+...,$$

$$(I,3n):v_{0}(3n)=\mp \frac{6\sqrt{3n}}{c_{1}\sqrt{f_{n-1}}}+...,$$

$$(III,3n):\pm 2\sqrt{3n}[-3\pm
2\sqrt{3n}-c_{1}\sqrt{f_{n-1}}+c_{1}\sqrt{f_{n}}]=
\pm6\sqrt{3}n^{3/2}+...\Rightarrow$$

$$\pm\sqrt{3n}c_{1}(\sqrt{p_{n}}-1)\sqrt{f_{n-1}}=\pm6\sqrt{3}n^{3/2}+...$$
\noindent
The left-hand and the right-hand parts of the last equation have different
orders, whence we get the claim.\par

Proof of Lemma 1.3.\par

Let $v_{0}(n)=u_{0}(n)n+...$.
$$(I,n):[u_{0}(n)-6]^{2}n=u_{0}(n)v_{1}(n-1)v_{1}(n)+...$$
\noindent
The following variants arise:\par

1)$u_{0}(n)\neq 6$\par

Let $v_{1}(n)=u_{1}(n)n^{x}+...\Rightarrow
v_{1}(n-1)=u_{1}^{1}(n)n^{1-x}+...$.  If $x\neq 1/2$, then we get
contradiction with the equation (II) $\Rightarrow v_{1}(n)=O(\sqrt{n}),\
v_{0}(n)=O(n)$. Analysing the formulae (1.4), we get the claim. \par

2)Let $u_{0}(n)\equiv 6$\par

$$v_{0}(n)=6n+...,\ v_{1}(n)=o(\sqrt{n})=3+c_{1}(n)+c_{0}(n+1)$$

$$v_{0}(n)=c_{0}(n)c_{1}(n).\ Let\ c_{0}(n)\simn^{x},\ c_{1}(n)\simn^{1-x}
\Rightarrow x=1/2.$$
\end{document}